\documentclass[pdflatex,11pt,sn-mathphys-num]{wlscirep}
\usepackage[utf8]{inputenc}
\usepackage[T1]{fontenc}
\usepackage{xcolor,soul,framed} 
\colorlet{shadecolor}{yellow}
\DeclareGraphicsExtensions{.pdf,.jpeg,.png}

\usepackage{array}
\usepackage{mdwmath}
\usepackage{eqparbox}
\usepackage{url}
\usepackage{diagbox}
\usepackage{tabularx}
\usepackage{graphicx}
\usepackage{adjustbox}
\usepackage{rotating}
\usepackage{multirow}
\usepackage{subcaption}  

\usepackage{amssymb}

\usepackage{tikz}

\usepackage{xcolor}
\usepackage[linesnumbered,ruled,lined,longend]{algorithm2e}
\usepackage{amssymb}

\SetCommentSty{mycommfont}

\SetKwInput{KwInput}{Input}                
\SetKwInput{KwOutput}{Output} 
\providecommand{\keywords}[1]
{
  \small	
  \textbf{\textit{Keywords---}} #1
}

    \title{\textit{Eternal Sunshine of the Mechanical Mind}: The Irreconcilability of Machine Learning and the Right to be Forgotten}

\iftrue

\author[1,2]{Meem Arafat Manab}

\affil[1]{School of Law and Government, Dublin City University, Dublin, Ireland}
\affil[2]{School of Data and Sciences, BRAC University, Dhaka, Bangladesh}

\affil[1]{meem.manab2@mail.dcu.ie}
\affil[2]{meem.arafat@bracu.ac.bd}
\begin{document}

\fi

\begin{abstract}
As we keep rapidly advancing toward an era where artificial intelligence is a constant and normative experience for most of us, we must also be aware of what this vision and this progress entail. By first approximating neural connections and activities in computer circuits and then creating more and more sophisticated versions of this crude approximation, we are now facing an age to come where modern deep learning-based artificial intelligence systems can rightly be called thinking machines, and they are sometimes even lauded for their emergent behavior and black-box approaches. But as we create more powerful electronic brains, with billions of neural connections and parameters, can we guarantee that these mammoths built of artificial neurons will be able to forget the data that we store in them? If they are at some level like a brain, can the right to be forgotten still be protected while dealing with these AIs? The essential gap between machine learning and the RTBF is explored in this article, with a premonition of far-reaching conclusions if the gap is not bridged or reconciled any time soon. The core argument is that deep learning models, due to their structure and size, cannot be expected to forget or delete a data as it would be expected from a tabular database, and they should be treated more like a mechanical brain, albeit still in development.
\\[10pt]
\textbf{Keywords:}  \textit{Privacy, Data Protection, Right to be Forgotten, Machine Unlearning}

\end{abstract}


\keywords{Privacy, Data Protection, Right to be Forgotten, Machine Unlearning}

\flushbottom

\maketitle
%
%
\thispagestyle{empty}

\section{Preamble: The AI Boom}

In the 1990s, AI research had come to all but a halt, with the failure of early LISP machines\cite{lispm}, the budget cut from the US Department of Defence’s DARPA-funded Strategic Computing Initiative Program\cite{roland2002strategic}, and the frustratingly low performance of the rule-based expert systems\cite{harmonexpert} being just some of the sore points of that epoch. Fast forward three decades, and we are in the middle of an AI Boom, with artificial intelligence-enabled tools having become as ubiquitous as mobile phones and as useful as GPS, if not more. If anybody doubts how far we have come in the last twenty years alone, comparing Clippy from Microsoft Office circa 2000\cite{Maedche2016-cx} with Apple’s Siri, and then comparing Siri with today’s ChatGPT should suffice very well.

A large motivation for this progress in AI was the near-simultaneous availability of large datasets and mass-produced GPUs. Datasets like ImageNet\cite{imagenet}, which has more than 14 million images annotated and hierarchially arranged by human participants, made it possible to test and refine AI models, and the incredible performance in 2012 of AlexNet\cite{NIPS2012}, a convolutional neural network model, to classify ImageNet images could be considered one of the watershed moments in recent AI history. It outperformed its closest competitor by 10.8\%, with an error margin of only 15.3\% when allowed to predict five options instead of one while we consider the model correct as long as the correct response is among these top five (this is the reported top-5 accuracy). This error rate had been unimaginable up to that point. But what is more interesting is that one of the minds behind AlexNet was Geoffrey Hinton, who originally developed with two other colleagues the mathematics behind multi-layered neural networks\cite{backprop} in 1986. Truth is, the mathematics and the ideas were all there for decades (since at least the 1940s, according to Jürgen Schmidhuber\cite{schmidhuber2022annotated}, another progenitor of AI, who recalls Soviet mathematicians Vladimir Vapnik and Andrey Kolmogorov as indispensable precursors), but it was only when GPUs, i.e. graphical processing units, became readily available in the market that practical neural networks could be built. This was in no small part thanks to the video gaming industry which led to ever cheaper versions of GPUs, and what has followed is a quasi-exponential rise in share prices of graphics chip manufacturers like Nvidia as they diversified into AI\cite{nvidia}. Why did neural networks require GPUs in the first place? At some level, these models, just like biological brains, are built around the idea of parallel processing\cite{chetlur2014cudnn}, and CPUs cannot do parallel computations effectively, while GPUs, or in common parlance, graphics cards, are built to do precisely this because they were originally designed to render high-resolution video games in real-time on graphical displays, which is inherently a parallel process.

\section{The Legal Backdrop}
\label{Related_Work}
Now, what does any of this have to do with the Right to be Forgotten in particular? The Right to be Forgotten, in the case of the European Union, is actually of a rather limited scope. Often referred to as the Right to Erasure (of Data) as enshrined in GDPR’s Articles 17 and 19 and also Recital 65\cite{GDPR2016a}, it awards EU citizens the right to request the deletion of their personal data without any unreasonable delay. This can lead to the deletion of their records, for example, from search engine results, as immortalized in the European Court of Justice case “Google Spain v AEPD and Mario Costeja González” and its historic judgment from 2014\cite{ECLI2014}. There have been talks of how this deletion may not be as instantaneous as we want it to be, due to, for example, search engines keeping large amounts of data in multiple servers and so-called cache-s (pronounced like cash) to make the search process as fast and smooth as possible, as Fosch-Villaronga et al. pointed out in their seminal work Humans Forget, Machines Remember: Artificial Intelligence and the Right to Be Forgotten\cite{VILLARONGA2018304}. But ultimately, this line of reasoning still sees data as information kept in some table or some database, and with machine learning, that might not quite be the case.

\section{Welcome to the Machine}
\label{background}

We should stress here that most AI technologies that have been either changing the world or making the headlines in the last decade, be they the facial recognition systems, deep fakes, or large language models like GPT-3 and Bard, are examples of machine learning, or more specifically its sub-branch called deep learning, where large amount of data is fed into an algorithm of layers of neural networks which identifies patterns in the existing data, and then either categorises or generates data using these patterns. These deep learning models employ internal parameters, known as weights and biases, for their tasks of pattern recognition and data generation. The parameter values are derived from some ‘training’ data provided to the algorithm and then used in the mathematical calculations for classification or creation of new data. Statistical in nature, the size of these models, i.e. the number of their parameters, has generally been smaller than the data they are trained with. Take GPT-4, as a reference. According to some sources, it has 1.7 trillion parameters and it was trained on 13 trillion tokens (small chunks of words and sentences)\cite{gpt4}, while GPT-3 has 175 billion parameters\cite{gpt3}. The information of these tokens is then remembered using the parameter values, so each token might be remembered using a handful of parameters, and each parameter could also be responsible for remembering some different tokens. Parameters, in this case, are doing what neurons do in our brains. If we want GPT-4 to “forget” the sentence: “Harry Potter can be treated as a fan-fiction of Lord of the Rings”, assuming each word in this sentence is a token, deleting one or two parameter values from the 1.7 trillion should be enough.

The problem is that we do not know which parameter in reality captures which information. Some scholars have rightly utilized neural networks for encryption, rather than compression\cite{Patel2021-nx}. Even if a machine learning model is not smaller than the data it has been trained on, as is often the case for the likes of zero-shot\cite{Xian_2017_CVPR} and few-shot learning\cite{wang-few-shots} models, it is still quite impossible to reverse-engineer the model and say with certainty that these “neurons” or these parameters are responsible for remembering this specific piece of data and this piece only. Just like an animal brain, trying to delete one piece of data might lead to a loss of information related to something else. We may not even know with great certainty which records would thus be affected since we do not know which pieces of information are correlated in the internal representation of the model. There have been recorded instances that deleting the information of one person can result in the deletion of information of other people with the same or similar names\cite{xu2023machine}. Researchers have been comparing neural networks to a hypothetical black box\cite{savage} since their early days\cite{castel}, and if these networks are not simply a collection of data, but rather a black box that consumes data and performs intelligent tasks based on them, deleting the data should be very difficult. 

And it indeed is. Most machine unlearning research (unlearning is the process by which a machine learning model forgets its training data) shows an accuracy far below\cite{xu2} what would be acceptable on the legal grounds of protecting fundamental rights. Turning back a century, cases like Melvin v. Reid (1931) and Sidis v. FR Publishing Corp. (1940)\cite{bates} from the United States first brought legal attention to the problem of being forgotten from the public eye. It was in relation to having personal information published without consent in films (Reid), books, or newspapers (Sidis) that these plaintiffs sought justice. Questions about freedom of speech or freedom of expression or perhaps undue censorship can then be brought against this right to claim relative obscurity. As long as things were confined to tabular data stored in some Spreadsheet file and then being Googled by strangers, the parallels had been there. But when we started to upload data into digital or electronic brains, things became convincingly more difficult. How do you make sure that ChatGPT will not remember you? What if the machine starts to lie? It has already been documented that large language models can have their memory distorted when tasked with learning new data\cite{ye2023cognitive}, or may even appear to hallucinate\cite{hall}. Faced with the challenge to forget some data, the distortion in their memory can be catastrophic. 

\section{Path to Reconciliation}

There is a rather obvious way toward resolving this conundrum, which is re-training the machine learning models from scratch, known in the literature as exact unlearning\cite{chundawat2023zero}. The trade-off here is even more disproportionate, as sources believe that it took 34 days for 1024 servers\cite{energy} (that is 1024 high-quality CPUs and GPUs) to train GPT-4. Now, if we wanted the personal data of the authors of this article, for example, to be deleted from GPT-4, retraining it would, in addition to time and server, require around 50 GWh of energy\cite{thinair}, which is enough to power 44 million households in the US for one year (that is one-third of the United States, and 24 times the size of Ireland). The other option, to use the currently available machine unlearning algorithms, would mean deleting it with a non-zero error margin\cite{sekhari2021remember}, which means that there will always remain a tiny but real chance that our data might not be deleted at all after our request has been processed. There is also a lot of ongoing research into differential privacy, which essentially means adding noise to all personal data for maximum anonymization. But it is still in an ongoing phase, and no large scale industrial development has been deployed yet. 

What if neural networks stopped being black boxes? If we could look at a machine learning model and tell which data it was trained on, and which parameters correspond to which pieces of data, then deleting corresponding parameters would be easy and straightforward, but a lot of information would then become compromised. For example, Meta built its own large language model LLaMA\cite{touvron2023llama} in 2023, and its weight parameters (in four versions, from 7 to 65 billion) have been freely available to download. If the data could be derived backward from these parameter values, we would have access to an abundance of text data that may or may not include people’s personal data as well.

There are also issues previously discussed in other publications, such as, for example, how these machine learning models, especially the large language models, store their data in intermediary servers, again for the sake of speed. This approach, usually referred to as “lazy evaluation”\cite{lazy} in computer science literature, can lead to retaining some data although it has been deleted officially. But even without lazy evaluation and caching, machine learning and the right to be forgotten are fundamentally at odds with each other. With machine learning, we are ever more approaching the creation of an artificial brain, often panderingly called AGI (Artificial General Intelligence), and the question now morphs into this: should we be able to delete data from a brain? What if it affects the workings of the rest of that brain? As we moved from rule-based to probabilistic artificial intelligence, we also moved from tabular databases that looked like gigantic Spreadsheets to neural networks, and deletion would now mean more than selecting cells in a table and then deleting the corresponding records. 

\section{Coda: The Digital Amnesiac}

None of the versions of the AI act\cite{aiact} currently available on the internet address this gap between machine learning and the right to erasure, although they go to a good length to explain what machine learning is, and some previous legal literature had sparingly acknowledged the disparity\cite{weissinger2021ai} before. Meanwhile, at least some computational researchers have argued that machine unlearning is only achievable mathematically, not practically\cite{thudi2022necessity}. Several potentially prohibitive solutions are still available for implementation, such as barring all training of machine learning, essentially deep learning, models using biometric or personal data until a foolproof and guaranteed machine unlearning method is proven to exist for that particular model. Another way could be to make AI tools like ChatGPT respond that they are unauthorized to provide a particular individual's personal data, rather than delete this data from within the underlying model itself. But then arises the risk of somebody deceiving the model (be it GPT-4 or Bard) to divulge the information\cite{floridi2023machine}. Or perhaps, the European Court of Justice could redefine or explain further what the Right to be Forgotten should mean in the age of AI. Geoffrey Hinton, arguably the father of neural networks and whom we mentioned at the beginning of this article, prescribed in 2022 to create specific-purpose computers for next-generation AI models\cite{hinton2022forward}, effectively barring people from doing AI research at home and turning transfer of parameters meaningless, through what he calls “mortal computation”. But while that would eliminate the risk of transferring data via transferring parameter, what will happen to the question of deletion? The authors of this article believe that as machine learning models become more and more like an anatomical brain, both in terms of performance and structure, the more irreconcilable they will become with the Right to be Forgotten. Can we ask a brain to forget us, even if we created it ourselves? It would perhaps be wiser, instead of taking reconciliatory attempts, to never introduce ourselves to that brain in the first place. Otherwise, in addition to forgetting one person named Clementine K.\cite{eternal}, the mechanical mind\cite{crane2015mechanical} could end up forgetting songs and fruits by the same name, with a digital amnesiac instead of an artificial general intelligence standing before us.

\section*{Data Availability}
No data has been generated or analyzed during this study at any phase.

\bibliography{eternal}

\end{document}